\documentclass{elsart}

\usepackage{amsmath}
\usepackage{amssymb}
\usepackage{graphicx}
\usepackage{dcolumn}
\usepackage{bm}

\def\be{\begin{eqnarray}}
\def\ee{\end{eqnarray}}
\def\ba{\begin{array}}
\def\ea{\end{array}}
\def\nn{\nonumber}
\begin{document}

\begin{frontmatter}

\title{A self-consistent microscopic model of Coulomb interaction in a
bilayer system as an origin of Drag Effect Phenomenon}

\author[l1,4]{ K. G\"uven},
\author[l2]{A. Siddiki},
\ead{siddiki@fkf.mpg.de}
\author[l1]{P. M. Krishna} and
\author[l1,3]{T. Hakio\u{g}lu }

\address[l1]{Physics Department, Bilkent University, Ankara, 06800 Turkey}
\address[4]{Nanotechnology Research Center, Bilkent University, Ankara, 06800 Turkey}
\address[l2]{Ludwigs-Maximillians University, Physics Dept., LS von Delft, D-80333 Munich, Germany}
\address[3]{UNAM Material Science and Nanotechnology Research
Institute, Bilkent University, Ankara, 06800 Turkey}

\begin{abstract}
In this work we implement the self-consistent Thomas-Fermi
model that also incorporates a local conductivity model to an electron-electron
bilayer system, in order to describe novel magneto-transport properties such as
the Drag Phenomenon. The model can successfully account for the poor screening of the potential
within the incompressible strips and its impact on the interlayer Coulomb interaction. An externally
applied current in the active layer results in the tilting of the Landau levels and built-up of a
Hall potential across the layer, which, in turn, induces a tilted potential profile in the passive
layer as well. We investigate the effect of the current intensity, temperature, magnetic field, and
unequal density of layers on the self-consistent density and potential profiles of the bilayer system.
\end{abstract}
\end{frontmatter}
%

\section{Introduction}
In a two-dimensional bilayer system of charged particles where the
separation of the layers are large enough to prevent tunnelling,
an external current applied in one (active) layer can drag the
charges in the adjacent (passive) via the inter-layer Coulomb
interaction, and thus, induce a current. This phenomenon, known as
the Coulomb Drag (CD) effect, which was proposed
theoretically~\cite{Price83:750}, and investigated
experimentally~\cite{Gramila91:1216}. The Coulomb drag is of
fundamental interest, since it can be utilized as a probing tool
to investigate the screened interaction in mesoscopic transport
systems. A quantitative measure of the Coulomb drag is the
transresistivity $\rho_{21} = E_2 / J_1$ where $J_1$ is the
applied current in the active layer and $E_2$ is the induced
electric field in the passive layer.

The presence of perpendicularly applied magnetic field enriches
the physics of the CD effect by introducing Landau levels. The
electrostatic and transport properties of the bilayer system were
investigated first within the independent electron
picture~\cite{Jauho93:4420} and recently within a self-consistent
screening scheme~\cite{siddikikraus:05,Bilayersiddiki06:}. The
main outcomes of the self-consistent scheme are (i) the formation
of incompressible strips in the active layer at integer Landau
filling factor values, which affects the density and potential
profile in the passive layer, and (ii) the tilting of the total
potential within the active layer in the out-of-linear response
(OLR) regime, induced by a large external
current~\cite{Guven03:115327}.

In this work, we combine these effects to formulate the interaction between
the active and passive layers within the OLR regime, in
order to provide a self-consistent description of the CD effect.
We investigate the effect of the current intensity, temperature
and density mismatch between the layers on the inter-layer Coulomb
interaction in the OLR regime.
\section{Model, Results and Discussion}
The electron bilayer system in the present work consists of
two $\rm{GaAs}$ quantum wells (having a width of $2d\sim3$ $\mu$m) that are
sandwiched between two silicon doped
thick $\rm{(AlGa)As}$ layers. This structure is grown on top of a $\rm{GaAs}$
substrate. The layers are translationally invariant.
The electron density of the top layer is manipulated by
a top gate. The electron layers are symmetric (with respect to the
growth direction) and ($h\sim$) $10-30$ nm apart separated by an
undoped $\rm{(AlGa)As}$ layer. Such a bilayer system is known to
be electronically decoupled and can be represented by two
different electrochemical potentials. We note that the
electron tunnelling between the layers is not possible. Both of
the electron channels are formed in the interval $-b<x<b$, where
$|d-b|$ sets the depletion length.

The total electrostatic potential of an electron on the line
$(x,z)$ due to a line-charge at $(x_{0},z_{0})$ is given
by~\cite{Bilayersiddiki06:} \be V(x,z)= V_{bg}(x,z) + V_H(x,z),
\label{eq:Vvon_x}\ee and \be V_H(x,z)= \frac{2e^2}{\bar{\kappa}}
\int_{x_{l}}^{x_{r}} dx_{0} K(x,x_{0},z,z_{0}) n_{\rm
el}(x_{0},z_{0}), \label{eq:V_Hartree} \ee where $-e$ is the
charge of an electron, $\bar{\kappa}$ an average background
dielectric constant, and the kernel $K(x,x_{0},z,z_{0})$ solves
Poisson's equation under the relevant boundary conditions given by
\be \label{perp-bilayer} K(x,x_{0},z,z_{0})=-\ln\!
\left(\frac{\cos ^2 \frac{\pi}{4d}(x+x_{0}) + \gamma ^2}{ \sin ^2
\frac{\pi}{4d} (x-x_{0}) + \gamma ^2}\right) \ee where the $z-$
dependence is given by $\gamma=\sinh (\pi |z-z_{0}|/4d)$. The
confining (background) potential is obtained by inserting a
constant number density ($n_{0}$) of the background charges into
Eq.~(\ref{eq:V_Hartree}). The gate can be
described by an induced charge distribution $n_{g}(x)$, that is residing
on the surface. In order to obtain a flat (gate) potential profile
in the bulk we choose the induced charge distribution as
$n_{g}(x)=n_{g}^{0}(1+\alpha(x/d)^2)$ where $n_{g}^{0}$ determines
the strength of the gate potential whereas $\alpha (=0.7)$ gives
the slope of the induced charge distribution. With such treatment
of the gate, the average electron densities can be modified while keeping
the depletion length constant. The Hartree potential
can be calculated from Eq.(\ref{eq:V_Hartree}) for the top (bottom) layer
yielding a total potential  \be V(x,z)=
-V^{T}_{bg}(x,z)-V^{B}_{bg}(x,z)-V_{g}(x,z)\nn \ee \be
+V_{H}^{T}(x,z)+V_{H}^{B}(x,z) .\label{int1d}\ee
The electron densities are calculated within the TFA: \be
\label{eq:TFA}n^{T,B}_{el}(x)=\int
dED(E)f([E+V(x,z_{T,B})-\mu^{*}_{T,B}]), \ee with $D(E)$ is the
Landau density of states (DOS), $f(E)=1/[\exp(E/k_BT)+1]$ is the
Fermi function, $\mu^{*}_{T,B}$ is the chemical potential (being
constant in the equilibrium state) and $z_{T,B}$ is the position of
top (bottom) layer, respectively. We employ the Gaussian
broadened Landau DOS~\cite{Guven03:115327}. The self-consistent numerical
calculations start with obtaining the confining potential created by its own donors for each
layer and the electron densities and the intralayer and interlayer Hartree
potentials~\cite{Siddiki03:125315} at zero temperature and zero magnetic field.
The energies are scaled by the average pinch-off energy $E_0=2\pi e^2 n_0 d/\bar{\kappa}$. The
lengths are scaled by the screening length $a_{0}$
($=a^{*}_{B}/2$) expressed in terms of effective Bohr radius,
$a^{*}_{B}=\bar{\kappa}\hbar^{2}/(me^{2})$. The electron density
and the electrostatic potential for finite temperature and magnetic field can be calculated
self-consistently by the above scheme within the TFA.

\begin{figure}
{\centering
\includegraphics[width=1.0\linewidth]{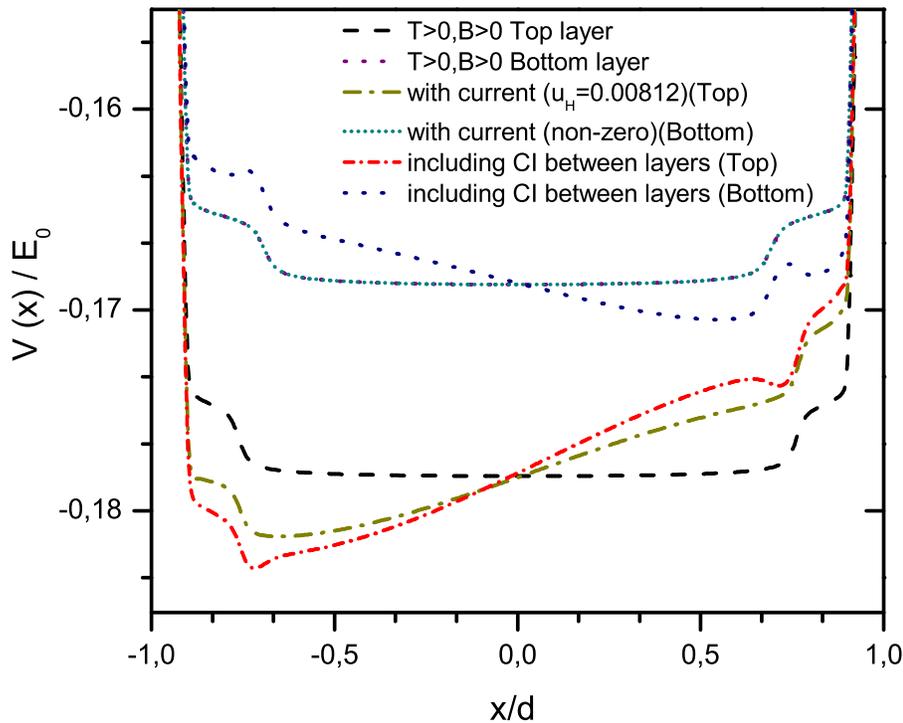}
%
\caption{ \label{fig:fig1}The calculated potential profiles as a
function of lateral coordinate, for zero and finite current applied
to the top layer. The gate potential is chosen such that the
resulting density mismatch is at the order of few percents, namely
$V_g/E_0=0.05$. Calculations are done at $k_BT/E_0=5.5\times10^{-3}$ for
$n_0=4\times10^{11}$ cm$^{-2}$ and interlayer spacing of $h=12$
nm. The $B$ field is selected such that
$\hbar\omega_c/E_0=5.5\times10^{-4}$ assuming a fixed depletion
length $b/d=0.9$}}
\end{figure}

The crucial and distinguishing feature of the approach we
implement is that we neither rely on a one-electron description
nor use any phenomenological parameters to account for electron
interactions, both for inter- and intra-layer. In the next step,
we drive a relatively high current from the active (here, top)
layer and calculate the effect of this on the passive layer. We
use a local version of the Ohm's law to describe current densities
employing a Gaussian broadened DOS defining the conductivity
model~\cite{Guven03:115327,siddiki2004}. We assume a local thermal
equilibrium, where local quantities vary slowly on the
quantum mechanical length scales. In the presence of current, it is
well known that the Landau levels are tilted due to the
Hall potential developed~\cite{Guven03:115327}, therefore the
electrochemical potential $\mu^{*}_{T,B}$, becomes position
dependent and thus must be included within the self-consistent scheme.
This brings two other self-consistent loops:
(i) the re-calculation of the $\mu^{*}_{T,B}(x)$ in each layer;
(ii) the response of the other layer to this modification. In this
work, we also include this tilting to investigate its effect on
the potential (see Fig~\ref{fig:fig1}) and the density
profile of the passive layer.

Here, we aim to show the effect of the combination of the inter-
and the intra-layer Coulomb interactions together with the effect
of external current driven in at least one of the layers. If one
drives a current perpendicular to the applied magnetic field a
Hall potential develops in the transverse direction to the
current, i.e. Hall potential, resulting from the Lorentz force. If
the system is all compressible, the Hall potential is linear in
$x$, as one would expect to observe in a normal metal. Here we
observe a similar behavior for the top layer, depicted in
Fig.\ref{fig:fig1}. On the other hand the response of the second
layer is as expected in the first glance, since, if the top layer
has a tilted electrochemical potential in one direction the other
should be the opposite. In Fig.~\ref{fig:fig1}, we show the
self-consistently calculated potential profiles of both layers.
For zero bias the top layer (broken black line) and the bottom
layer (dotted cyan line) are symmetric in $x$. However, for a
large current intensity ($U_H$, see Ref.~\cite{Guven03:115327}) at
the top layer to induce an electrochemical potential of this layer
(red short dashed-dotted lines) a tilt is induced at the bottom
layer with an opposite slope (dark blue dotted lines). More
interestingly, if a finite current is driven from the bottom layer
with same current direction of the top layer, the tilting observed
is slightly suppressed (brown dashed-dotted line).

In summary, by exploiting the smooth variation of the external
potential we have explicitly calculated the Coulomb interaction in
and between the bilayer quantum Hall samples in the presence of a
strong current driven in one layer. Our preliminary results
indicate that, a strong Coulomb coupling induces modifications in
the potential landscape of both layers, which clearly points that
a self-consistent microscopic model is necessary to grasp the
underlying physics of the Coulomb drag effect observed in the
bilayer systems.

The authors acknowledge the support of the Marmaris Institute of
Theoretical and Applied Physics (ITAP), TUBITAK grant 105T110,
SFB631 and DIP.

\end{document}